\documentclass[prd,nofootinbib,showpacs,twocolumn]{revtex4-1}
\usepackage{color}
\usepackage{bm}
\usepackage{graphicx}
\usepackage{amsmath}
\usepackage{amssymb}
\usepackage{enumitem}
\usepackage{hyperref}
\usepackage{subfigure}
\usepackage{array}
\usepackage[english]{babel}
\usepackage{tensor}
\usepackage{comment}
\usepackage[normalem]{ulem}
\usepackage[dvipsnames]{xcolor}

\def\be{\begin{equation}}
\def\ee{\end{equation}}
\def\ba{\begin{eqnarray}}
\def\ea{\end{eqnarray}}
\def\l{\left}
\def\r{\right}

\allowdisplaybreaks

\begin{document}

\title{Signatures of $f(Q)$-gravity in cosmology}

\author{Noemi Frusciante}
\affiliation{
\smallskip
 Instituto de Astrofis\'ica e Ci\^{e}ncias do Espa\c{c}o, Faculdade de Ci\^{e}ncias da Universidade de Lisboa, Edificio C8, Campo Grande, P-1749016, Lisboa, Portugal }

\begin{abstract}
We investigate the impact on cosmological observables  of  $f(Q)$-gravity, a specific class of modified gravity models in which gravity is   described by the non-metricity scalar, $Q$. In particular we focus on a specific model which is indistinguishable from the  $\Lambda$-cold-dark-matter ($\Lambda$CDM) model at the background level, while showing peculiar and measurable signatures at linear perturbation level.  These are attributed to a time-dependent Planck mass and are regulated by a single dimensionless parameter, $\alpha$. In comparison to the $\Lambda$CDM  model, we find for positive values of $\alpha$ a suppressed  matter power spectrum and lensing effect on the Cosmic Microwave Background radiation (CMB) angular power spectrum  and an enhanced integrated-Sachs-Wolfe tail of CMB temperature anisotropies. The opposite behaviors are present when the $\alpha$ parameter is negative. 
 We also investigate the modified Gravitational Waves (GWs) propagation and show the prediction of the GWs luminosity
distance compared to the standard electromagnetic one. Finally, we infer the accuracy on the free parameter of the model with standard sirens at future  GWs detectors.

\end{abstract}


\date{\today}

\maketitle

\section{Introduction}

The late-time cosmic acceleration has been confirmed by different cosmological observations~\cite{Riess:1998cb,Perlmutter:1998np,Betoule:2014frx,Spergel:2003cb,Ade:2015xua,Aghanim:2015xee,Eisenstein:2005su,Beutler:2011hx}. 
Within General Relativity (GR), it is the cosmological constant $\Lambda$ to give rise to the observed  acceleration of the Universe. In this picture the resulting standard cosmological model ($\Lambda$CDM), besides providing an accurate description of the Universe,  comes along with some theoretical problems~\cite{Joyce:2014kja} and mild observational tensions, i.e. on  the measurements of the value of the Hubble constant $H_0$~\cite{Adam:2015rua,Aghanim:2018eyx,Riess:2011yx,Riess:2016jrr,Riess:2019cxk,Delubac:2014aqe} and the amplitude of the matter power spectrum at present time $\sigma_8$~\cite{Hildebrandt:2016iqg,deJong:2015wca,Kuijken:2015vca,Conti:2016gav,Joudaki:2019pmv} from different surveys. These might signal the necessity of looking for new physics beyond the standard model. 

Several modified gravity (MG) proposals have been considered which modify the gravitational interaction on cosmological scales. Following the GR construction,  most of them have null non-metricity and torsion~\cite{Nojiri:2010wj,Lue:2004rj,Copeland:2006wr,Silvestri:2009hh,Capozziello:2011et,Clifton:2011jh,Tsujikawa:2010zza,Joyce:2014kja,deRham:2014zqa,Heisenberg:2014rta,Koyama:2015vza,Nojiri:2017ncd,Ferreira:2019xrr,Kobayashi:2019hrl,Frusciante:2019xia}.  Alternatively one can construct theories of gravity built from the scalars associated to torsion ($T$) and  non-metricity ($Q$).  While the actions  $\int d^4x\sqrt{-g}\,T$ and $\int d^4x\sqrt{-g}\,Q$ are equivalent to GR in flat space~\cite{BeltranJimenez:2019tjy}, their generalizations with $f(T)$~\cite{Li:2010cg,Wu:2010mn,Wu:2010xk,Cai:2015emx,Benetti:2020hxp} and $f(Q)$~\cite{Nester:1998mp,Dialektopoulos:2019mtr,Jimenez:2019ovq,Lu:2019hra,Lu:2019hra,Bajardi:2020fxh} can be ascribed in the class of MG models. In the following we focus on $f(Q)$-gravity which introduces at least two additional scalar modes. These disappear around maximally symmetric backgrounds causing strong coupling problems. However, while the $f(T)$-gravity models suffer from  strong coupling problems when considering perturbations around a Friedmann-Lema{\^i}tre-Robertson-Walker (FLRW) background~\cite{Golovnev:2018wbh},  these are absent in the case of $f(Q)$-gravity~\cite{Jimenez:2019ovq}. In $f(Q)$-gravity the main equations for the linear perturbations for scalar, vector and tensor modes and the matter density perturbation have been derived~\cite{Jimenez:2019ovq}. From this study, modifications in the evolution of the gravitational potentials and tensor propagation equations  emerge with respect to the $\Lambda$CDM model, thus making it worth to be further investigated by looking at the signatures these modifications leave on cosmological observables. Furthermore constraints on the deviations of the $f(Q)$-gravity from the $\Lambda$CDM background have been performed using different observational probes for several parameterizations of the $f(Q)$ function in terms  of redshift, $z$, i.e. $f(z)$~\cite{Lazkoz:2019sjl} and at linear perturbation level, the modification in the evolution of the matter density perturbation has been tested against  redshift space distortion (RSD) data~\cite{Barros:2020bgg}.

 In this work we focus on the $f(Q)$ model which shares the same background evolution as in $\Lambda$CDM, while leaving precise and measurable effects on cosmological observables. We perform a thorough analysis of its phenomenology at linear order  in perturbations, by comparing the predictions of the theory to the $\Lambda$CDM model for the temperature-temperature (TT) power spectrum, lensing potential auto-correlation power spectrum and matter power spectrum.  To this aim we implement the model in the public Einstein-Boltzmann code \texttt{MGCAMB}~\cite{Zhao:2008bn,Hojjati:2011ix,Zucca:2019xhg}. Finally we investigate the modification in the Gravitational waves (GWs) sector by presenting the GWs luminosity distance and we infer the accuracy on the free parameter of the model using previous forecasts from standard sirens~\cite{Belgacem:2018lbp,Belgacem:2019pkk} at the Einstein Telescope (ET)~\cite{Sathyaprakash:2012jk} and the Laser Interferometer Space Antenna (LISA)~\cite{Audley:2017drz}.

\section{$f(Q)$ model}

The action  of the $f(Q)$-gravity can be written as follows~\cite{BeltranJimenez:2017tkd}
\be
S=\int d^4x\sqrt{-g}\l[-\frac{1}{2}f(Q)+L_m\r]\,,
\ee
where $g$ is the determinant of the metric $g_{\mu\nu}$, $f(Q)$ is a general function of the non-metricity scalar $Q=-Q_{\alpha\mu\nu}P^{\alpha\mu\nu}$, with $Q_{\alpha\mu\nu}=\nabla_\alpha g_{\mu\nu}$ being the non-metricity tensor and $P^{\alpha}_{\phantom{\alpha}\mu\nu}=-L^{\alpha}_{\phantom{\alpha}\mu\nu}/2+\l(Q^\alpha-\tilde{Q}^\alpha\r)g_{\mu\nu}/4-\delta^\alpha_{(\mu}Q_{\nu)}/4$ where $Q_\alpha=g^{\mu\nu}Q_{\alpha\mu\nu}$, $\tilde{Q}_\alpha=g^{\mu\nu}Q_{\mu\alpha\nu}$ and $L^\alpha_{\phantom{\alpha}\mu\nu}=(Q^\alpha_{\phantom{\alpha}\mu\nu}-Q_{(\mu\nu)}^{\phantom{(\mu\nu)} \alpha})/2$. Finally, $L_m$ is the matter Lagrangian of standard matter fields.  The choice  $f=Q/8\pi G_N$, where $G_N$ is the Newtonian constant, reproduces the dynamics of GR. A particular class of $f(Q)$-theory that gives an expansion history on a FLRW background identical  to that of $\Lambda$CDM is~\cite{Jimenez:2019ovq}:
\be\label{model}
f=\frac{1}{8\pi G_N}\l(Q+M\sqrt{Q}\r)\,,
\ee
where $M$ is a constant. In the following we use the dimensionless parameter $\alpha\equiv M/H_0$, where $H_0$ is the present time value of the Hubble parameter $H(t)\equiv\frac{1}{a}\frac{da}{dt}$ and $a(t)$ is the scale factor.  Given the peculiar characteristic of the model at the background level,  the different values of $\alpha$ can thus only be depicted by analyzing the evolution of the linear perturbations. The model in Eq.~\eqref{model} will be the subject of the following analysis. Hereafter we will also make use of the redefinition: $f \rightarrow f/8\pi G_N$. 

\begin{figure}[t!]
\includegraphics[width=0.45\textwidth]{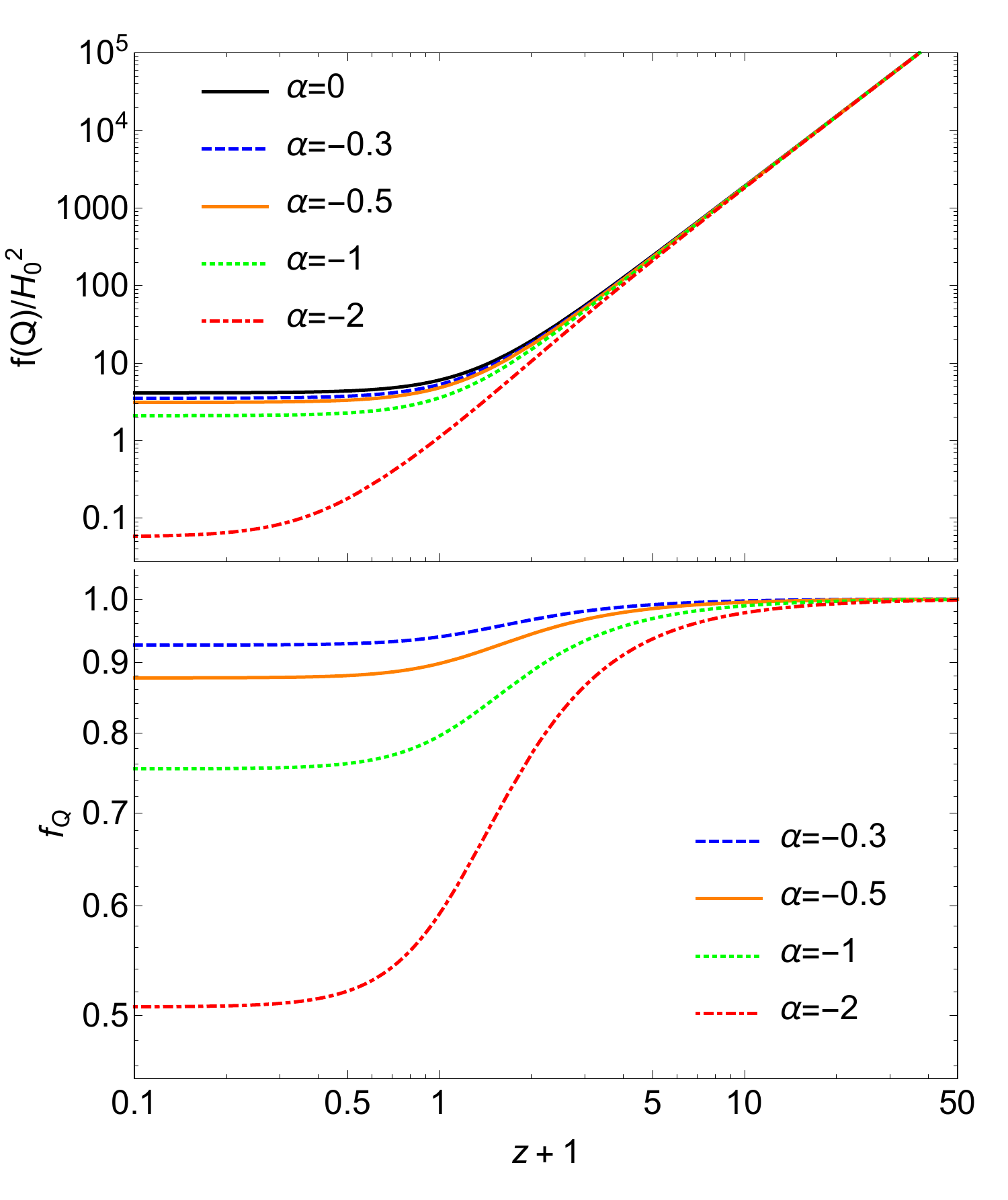}
\caption{Evolution of $f$ and $f_Q$ as a function of the redshift for negative values of the $\alpha$ parameter.}
\label{fig:behaviorsminus}
\end{figure}

\begin{figure}[t!]
\includegraphics[width=0.45\textwidth]{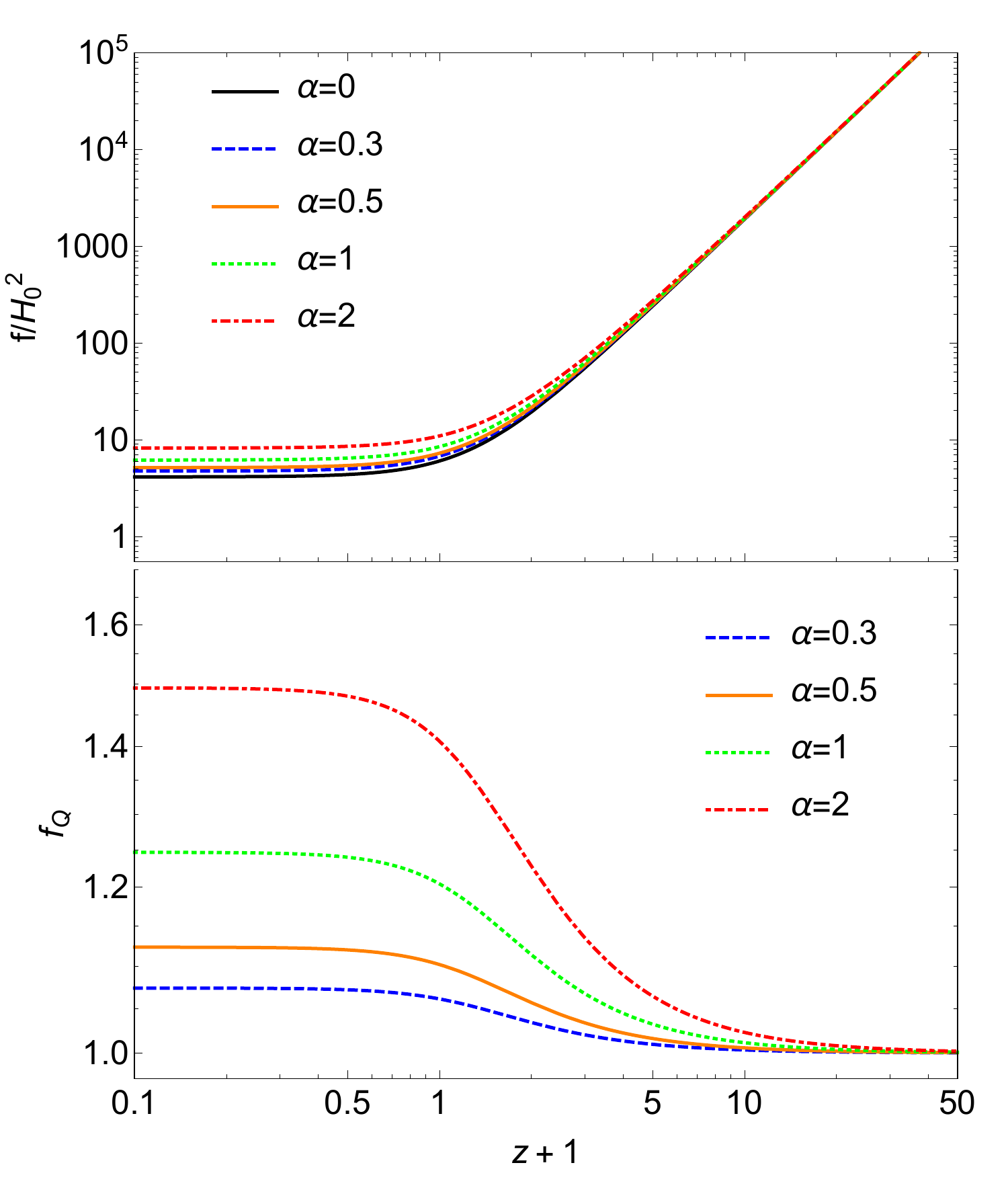}
\caption{Evolution of $f$ and $f_Q$ as a function of the redshift $z$ for positive values of the $\alpha$ parameter. }
\label{fig:behaviorsplus}
\end{figure}

Considering that the non-metricity scalar assumes the form  $Q=6H^2$ on a FLRW background, we show in Figs.~\ref{fig:behaviorsminus} and~\ref{fig:behaviorsplus} top panels the behaviors of the $f(Q)$ function given by the model in Eq.~\eqref{model} respectively for negative and positive values of the $\alpha$ parameter. The chosen values for $\alpha$ have the purpose to visualize and quantify the modifications. We also include as reference the $\alpha=0$ case corresponding to the GR limit. The values for the cosmological parameters used in  this paper are the Planck 2018 best fit values~\cite{Aghanim:2018eyx}.  Significant deviations from the GR behavior appear for smaller redshift ($z<4$)  and these differences start to vanish in the distant past. We can also note that while for positive values of $\alpha$,  $f(Q)$ is enhanced with respect to the GR limit, for negative ones it is suppressed. These different trends will impact the evolution of the linear perturbations. The derivative of $f(Q)$ with respect to the non-metricity scalar, $f_Q\equiv df/dQ$, indeed is identified to be the effective Planck mass~\cite{Jimenez:2019ovq} and as such it is expected to impact on the shape  of large scale observables. In the following we will investigate the physics of Microwave Background radiation (CMB) scalar angular power spectra, matter power spectrum and the implications of a modified propagation of GWs.

\begin{figure}[t!]
\includegraphics[width=0.45\textwidth]{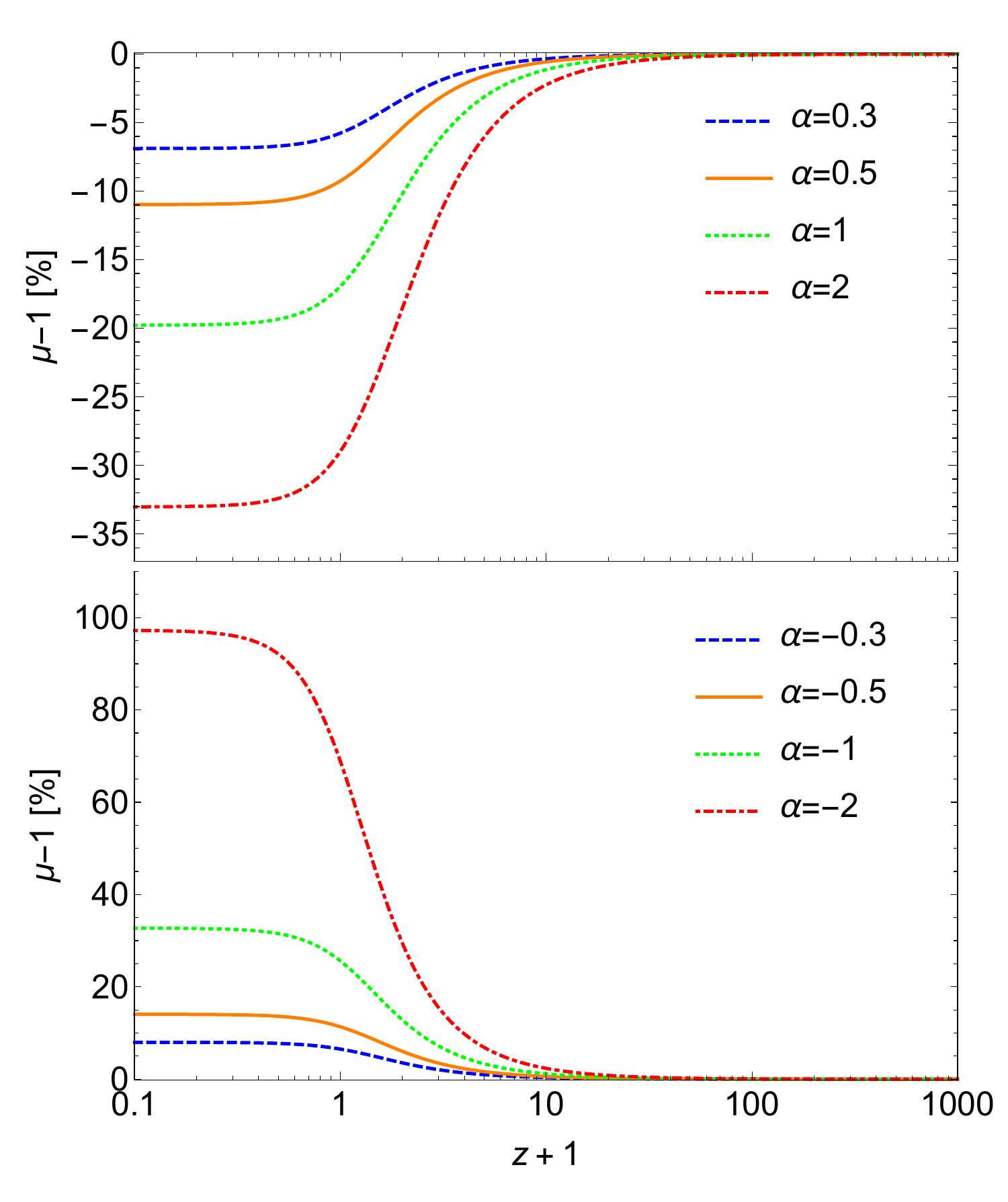}
\caption{Evolution of the phenomenological function $\mu$  as a function of the redshift $z$ for positive (top panel) and negative (bottom panel) values of the $\alpha$ parameter.  }
\label{fig:mu}
\end{figure}

\begin{figure}[t!]
\includegraphics[width=0.45\textwidth]{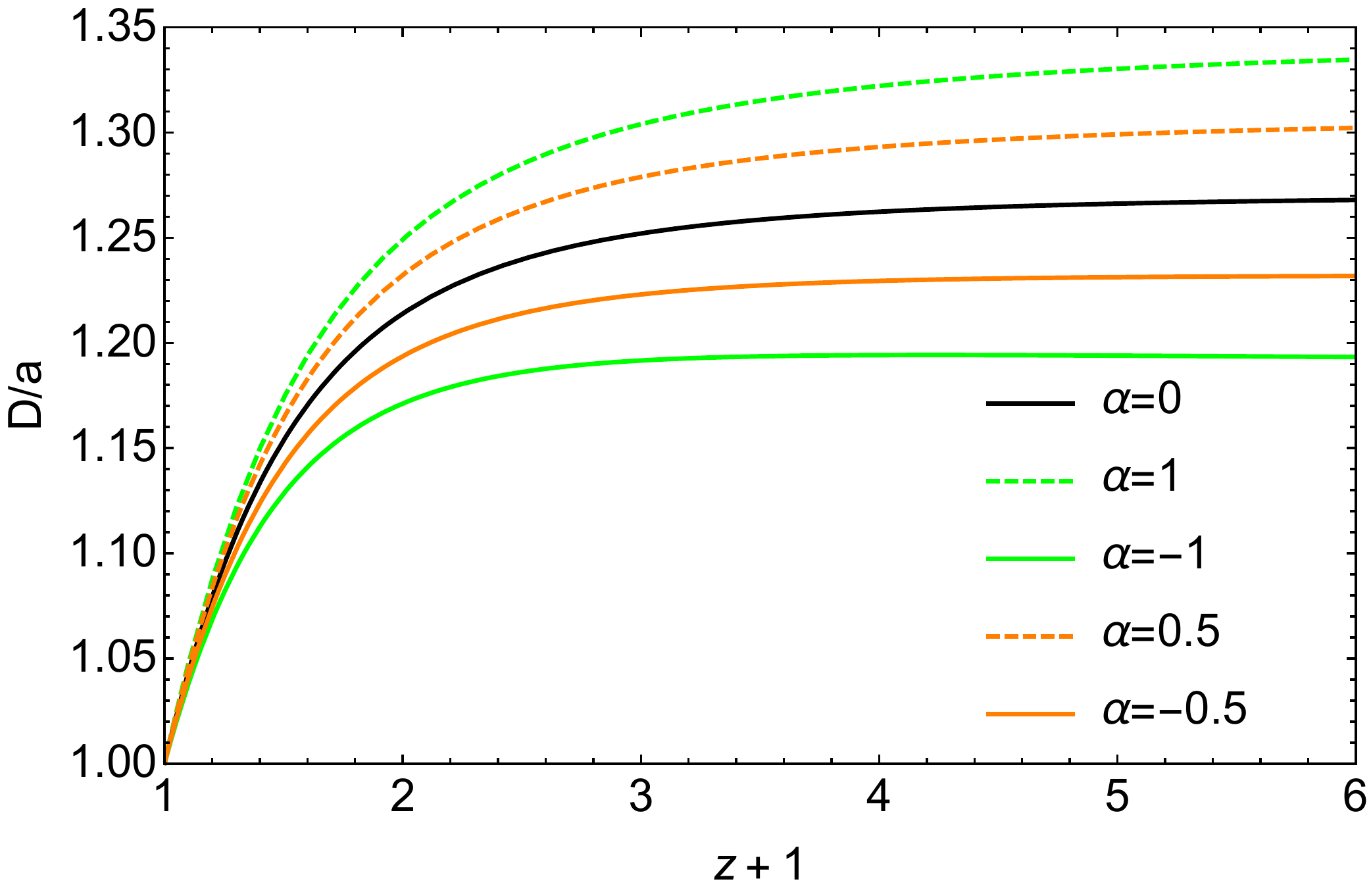}
\caption{Evolution of the linear growth factor $D$ normalized to unity today divided by the scale factor $a$ as a function of the redshift for different values of the $\alpha$ parameter. }
\label{fig:growth}
\end{figure}

\section{Effects on observations from the scalar sector}\label{Sec:effects}

Let us consider the perturbations of the metric in Newtonian gauge, $ds^2=-(1+2\Psi)dt^2+a^2(1-2\Phi)dx^2$, where $\Phi(t,x_i)$ and $\Psi(t,x_i)$ are the gravitational potentials. In the standard cosmological model, these two potentials are equal during the period of structures formation. This is no longer true when  modifications in the gravitational interaction are considered. In Fourier space the function $\eta(a,k)\equiv \Phi/\Psi$  defines the non-zero anisotropic stress, and the modifications in the Poisson equation are enclosed in the $\mu(a,k)$ function as:
\be
-k^2\Psi= 4 \pi G_Na^2\mu(a,k)\rho_m \delta_m\,,
\ee
where $\delta_m=\delta \rho_m/\rho_m$ is the density contrast, $\rho_m(t)$ is the background matter density and $\mu(a,k)$ defines the effective gravitational coupling. Moreover, a light deflection parameter $\Sigma(a,k)\equiv \mu(1+\eta)/2$ measures the deviation in the Weyl potential $(\Phi+\Psi)$. 

In the quasi-static approximation  for perturbations deep inside the Hubble radius for the $f(Q)$-gravity one has $\eta=1$ and $\mu=1/f_Q$~\cite{Jimenez:2019ovq}. In detail, for the model under consideration the latter turns out to be a function of time only and it reads:
\be\label{eq:mu}
\mu(a)= \frac{12 H}{12H+\sqrt{6}\alpha H_0} \,.
\ee

In Figs.~\ref{fig:behaviorsminus} and~\ref{fig:behaviorsplus} (bottom panels) we show the evolution of $f_Q$ as a function of the redshift. $f_Q$ being the effective Planck mass is always positive defined as expected. When $\alpha<0$, we have $f_Q<1$, thus  according to the evolution of $\mu$ the gravitational interaction is stronger than that in GR (see bottom panel in Fig.~\ref{fig:mu}). On the other hand when $\alpha>0$ the opposite holds (upper panel in Fig.~\ref{fig:mu}).  This aspect is quite interesting in light of RSD, galaxy clustering (GC) and  weak lensing (WL) data which independently  detected a lower growth rate of matter density perturbations than that predicted by the $\Lambda$CDM model \cite{Hildebrandt:2016iqg,deJong:2015wca,Kuijken:2015vca,Joudaki:2019pmv,Abbott:2017wau,Beutler:2013yhm,Samushia:2013yga,Macaulay:2013swa,Vikhlinin:2008ym}.  

In order to perform explorations of cosmological observables we have modified the public Einstein-Boltzmann code \texttt{MGCAMB} \footnote{ \texttt{MGCAMB} webpage: \url{https://github.com/sfu-cosmo/MGCAMB}}~\cite{Zhao:2008bn,Hojjati:2011ix,Zucca:2019xhg} which evolves the linear cosmological perturbations equations  taking into account the MG effects given by Eq.~\eqref{eq:mu}.

\begin{figure}[t!]
\includegraphics[width=0.45\textwidth]{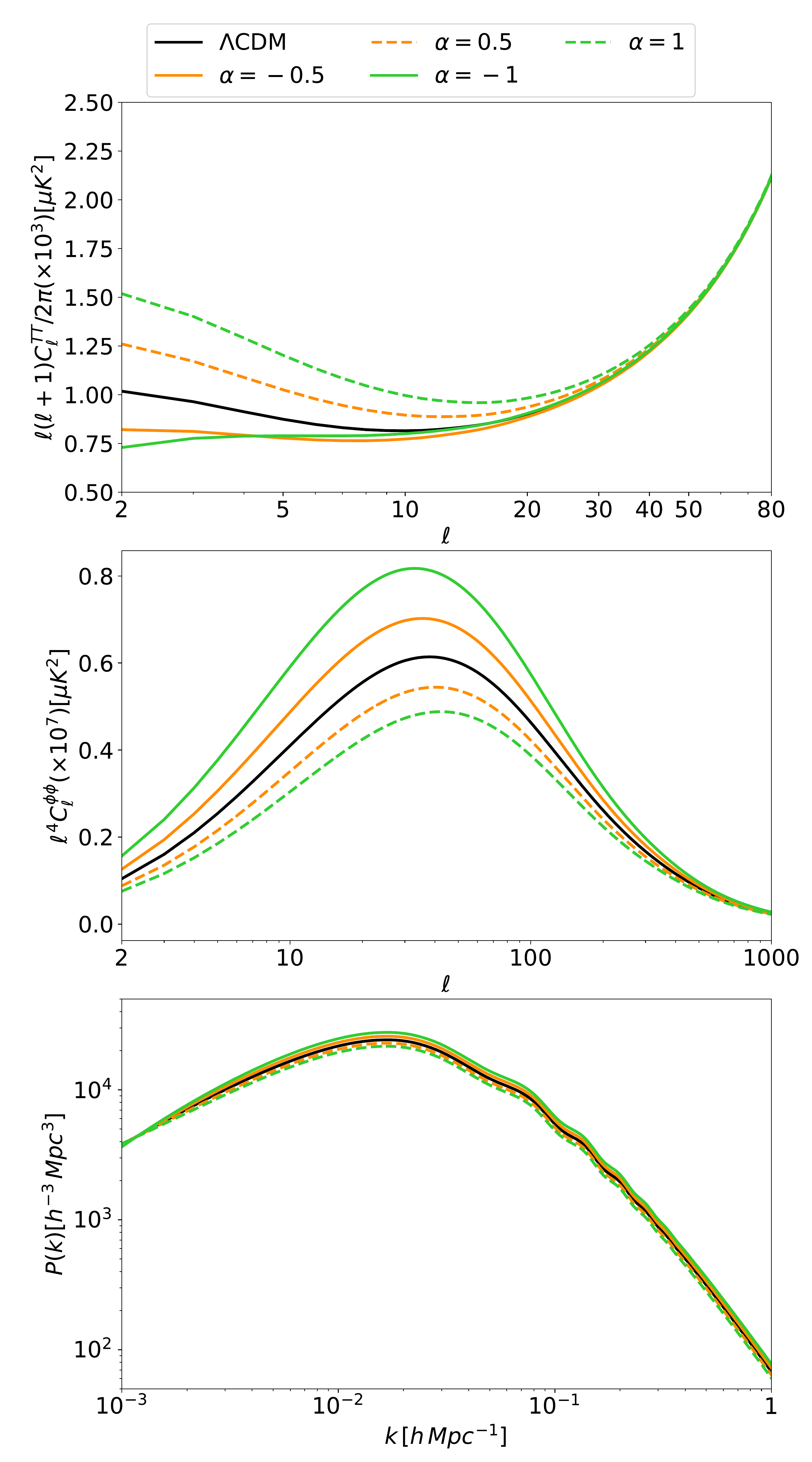}
\caption{Power spectra of different cosmological observables for different values of the $\alpha$ parameter and for the $\Lambda$CDM model. \textit{Top panel:} CMB temperature-temperature power spectra at low-$\ell$. \textit{Central panel:} lensing potential auto-correlation power spectra. \textit{Bottom panel:}  matter power spectra.}
\label{fig:spectra}
\end{figure}

The modification introduced by the model in Eq.~\eqref{model}  has three major effects on observations: 
\begin{enumerate} 
\item  \underline{It changes the growth of matter perturbations}. In modified gravity the  matter density perturbation $\delta_m$ obeys the linear equation:
\be
\ddot{\delta}_m+2H\dot{\delta}_m-4\pi G_N\mu \rho_m\delta_m=0\,,
\ee
which can be solved  by setting initial conditions in the matter dominated era. We show in Fig.~\ref{fig:growth} the redshift evolution of the growth factor $D(a)\equiv \delta_m(a)/\delta_m(a=1)$. As expected, the models with positive values of $\alpha$ and $\mu<1$ have a larger growth factor than $\Lambda$CDM as such, we predict a lower value for the amplitude
of the matter power spectrum at present time, $\sigma_8$, compared to $\Lambda$CDM,  assuming they share the same initial amplitude of primordial perturbations, $A_s$~\cite{Dolag:2003ui,DeBoni:2010nz,Pace:2013pea}.  Accordingly we observe   for positive values of $\alpha$ a  suppression of the growth of structures in the total matter power spectrum for $k\geq10^{-3}$ with respect to the $\Lambda$CDM one, as  depicted in the lower panel of Fig.~\ref{fig:spectra}. For the values considered the deviations are estimated to be in the range 5\%-10\%. When $\alpha$ is chosen to be negative the growth factor and the matter power spectrum show exact opposite characteristics, and a deviation of 15\% is found for $\alpha=-1$ in the matter power spectrum.  RSD data have been used to constrain the $\alpha$ and $\sigma_8$ parameters while keeping all the others fixed to the Planck 2018 best fit values. It is found that $\alpha=2.0331^{+3.8212}_{-1.9596}$ and $\sigma_8=0.8326^{+0.1386}_{-0.0630}$ at 1$\sigma$~\cite{Barros:2020bgg}. From their central values we can infer a preliminary estimation of the corresponding value for $A_s\simeq2.7\times10^{-9}$. 

\item \underline{It modifies the gravitational lensing}.  For the $f(Q)$-gravity the modification in the lensing gravitational potential $\phi_{\rm len}=(\Psi+\Phi)/2$ are associated to $\Sigma=\mu$. Thus for the specific case that we explore, the gravitational lensing is enhanced when $\mu>1$ and suppressed for $\mu<1$. Let us consider now the lensing potential auto-correlation power spectrum which using the line of sight integration method reads \cite{Lewis:2006fu}:
\begin{align}
\label{phiphispectrum}
C_\ell^{\phi\phi}=4\pi\int \frac{{\rm d}k}{k}\mathcal{P}(k)
\left[\int_0^{\chi_{\ast}}{\rm d}\chi\,
S_{\phi}(k;\tau)j_\ell(k\chi)\right]^2,
\end{align}
with $\mathcal{P}(k)=\Delta^2_\mathcal{R}(k)$ being the primordial power spectrum of curvature perturbations,  $j_{\ell}$ is the spherical Bessel function and 
\be\label{eq:lensing}
S_{\phi}(k;\tau)=2T_{\phi}(k;\tau_0-\chi)\left(\frac{\chi_{\ast}-\chi}{\chi_{\ast}\chi}\right)\,, 
\ee
where $T_{\phi}(k,\tau) = k\,\phi_{\rm len}$ is the transfer function, $\chi$ is the comoving 
distance  ($\chi_{\ast}$ corresponds to that to the last scattering surface),  with relation $\chi=\tau_0-\tau$ being $\tau$  the conformal time and  $\tau_0$ is its value today.
The modifications in the lensing potential $\phi_{\rm lens}$ discussed before, when included in the source term in Eq~\eqref{eq:lensing}, impact  on the lensing  power spectrum as shown in the central panel of Fig.~\ref{fig:spectra}. We note that for the negative values of $\alpha$ ($\mu>1$) the  lensing power spectrum is enhanced with respect to $\Lambda$CDM, while positive values correspond to a suppression of the lensing power ($\mu<1$). The larger deviations are for the higher values of $|\alpha|$. They reach the 50\% when $\alpha=-1$ and 25\% for $\alpha=1$.

\item \underline{It impacts the late-time Integrated Sachs-Wolfe} \underline{(ISW) effect}.   A modification in the time variation of the lensing potential is expected due to the presence of $\Sigma \neq 1$ for small $z$ which induces a late-time ISW effect. Let us consider the TT angular spectrum~\cite{Seljak:1996is}
\be
\label{TTspectrum}
C_\ell^{\rm TT}=
(4\pi)^2\int \frac{{\rm d}k}{k}~\mathcal P(k)\Big|\Delta_\ell^{\rm T}(k)\Big|^2\,,
\ee
with 
\be
\Delta_\ell^{\rm T}(k)=
\int_0^{\tau_0}{\rm d}\tau\,
e^{ik\tilde{\mu}(\tau-\tau_0)}S_{\rm T}(k,\tau)
j_\ell[k(\tau_0-\tau)]\,,
\ee
where $\tilde{\mu}$ is the angular separation, and 
$S_{\rm T}(k,\tau)$ is the radiation transfer function. The ISW contribution to  $S_{\rm T}(k,\tau)$  is given by
\begin{align}
\label{T_source}
S_{\rm T}(k,\tau) \sim \left(\frac{d \Psi}{d \tau}+\frac{d \Phi}{d \tau} \right)
e^{-\kappa}\,, 
\end{align}
where $\kappa$ is the optical depth. 
The late time variation of 
$\phi_{\rm lens}$ expected from $\mu$ thus induces the late-time ISW effect. 
We find that a suppression in the lensing power spectrum corresponds to  an enhancement of the amplitude of the low-$\ell$ TT power spectrum relative to $\Lambda$CDM, as shown in the top  panel of Fig.~\ref{fig:spectra}.  On  the contrary an enhancement in the  lensing power spectrum  results in a suppression of the large scale ISW tails.   As for the lensing case, the magnitude of the  deviations from $\Lambda$CDM depends on $|\alpha |$. We find that up to a 50\% deviation is present for $\alpha=1$ and  30\% for $\alpha=-1$.  The effect on the ISW tail  should be  then tightly constrained from the CMB data. We note that the realization of both a lower ISW tail and an enhanced matter power spectrum can be present in $f(R)$-gravity~\cite{Song:2007da} and  Galileon model~\cite{Peirone:2019aua} also.

\end{enumerate}

We showed that the $f(Q)$ model analyzed in this section leaves precise and measurable signatures on the CMB temperature anisotropies as well as on the matter power spectrum.   In particular some of them can be relevant to understand whether  the model can alleviate some tensions. For example, the model allows to realize weaker gravity than $\Lambda$CDM, and this can be very promising in light of the $\sigma_8$ tension. A preliminary result using only RSD data showed that indeed this can be the case~\cite{Barros:2020bgg}. However in the analysis only $\alpha$ and $\sigma_8$ are varied, while  all the other cosmological parameters are fixed. Thus a more general study is required which should include the variation of all other parameters and the use of several datasets as well. Moreover an enhancement in the lensing might accomodate the lensing excess in the CMB  Planck temperature data~\cite{Ade:2013zuv,Adam:2015rua,Aghanim:2018eyx}. Instead a suppressed ISW tail might accomodate better the CMB data over the standard cosmological scenario as it has been shown in  the Galileon Ghost Condensate model~\cite{Peirone:2019aua}. These features cannot be all present at the same time, thus a detailed   Markov chain Monte Carlo (MCMC) analysis  involving several  current observational data is needed.

\section{Gravitational waves luminosity distance}
\begin{figure}[t!]
\includegraphics[width=0.45\textwidth]{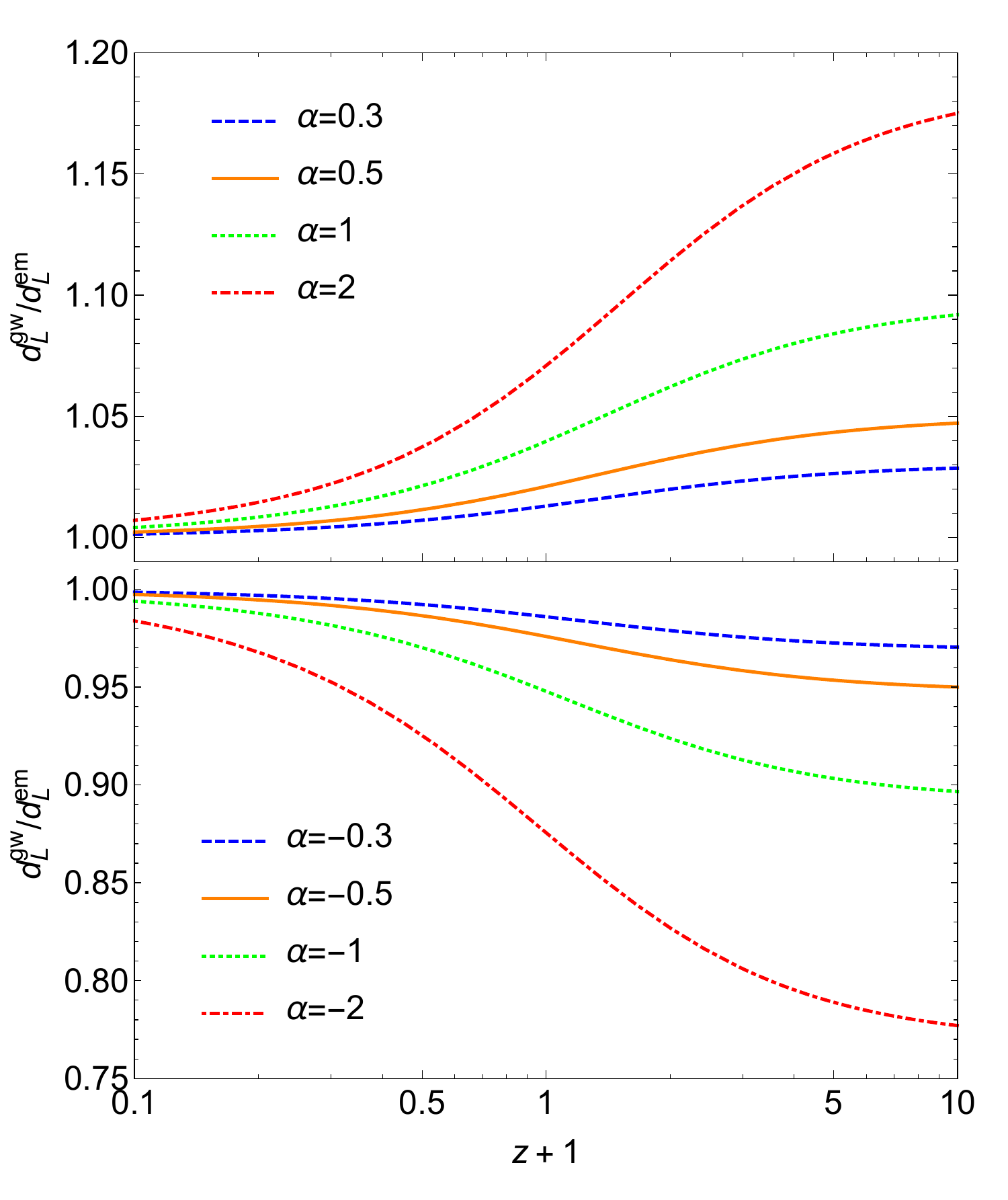}
\caption{Evolution of $d_L^{gw}/d_L^{em}$ as  function of the redshift for positive (top panel) and negative (bottom panel) values of the $\alpha$ parameter. }
\label{fig:Gws}
\end{figure}

The propagation of GWs modes obeys to the following  second order action in Fourier space~\cite{BeltranJimenez:2017tkd}:
\be
S= \frac{1}{2}\sum_\lambda\int d^3kd t \,a^3\, f_Q\l[(\dot{h}_{(\lambda)})^2-\frac{k^2}{a^2}h_{(\lambda)}^2 \r]\,,
\ee
where $h_{(\lambda)}$ are the two helicity modes of the metric tensor perturbation part. 
According to this action, the corresponding equation of propagation of GWs introduces a modification in the friction term which is  identified to be  
\be
 \delta(z)= \frac{d \ln \sqrt{f_Q}}{d(1+z)}\,.
 \ee
For $f(Q)$-gravity we can connect the friction term to the running of the effective Planck mass, $\alpha_M=-2\delta$, being $f_Q$ the effective Planck mass. A modified friction term affects the amplitude of  GWs such that the GWs luminosity distance, $d_L^{gw}$, is no longer equal to the standard electromagnetic  luminosity distance, $d_L^{em}$~\cite{Lombriser:2015sxa,Saltas:2014dha,Nishizawa:2017nef,Belgacem:2017ihm,Belgacem:2018lbp}. Indeed they are related by:
\be
d_L^{gw}(z)=d_L^{em}(z)\exp\l\{-\int^z_0 \frac{dz^\prime}{1+z^\prime}\delta (z^\prime)\r\}\,.
\ee
In Fig.~\ref{fig:Gws} we show the evolution of the ratio $d_L^{gw}/d_L^{em}$ as a function of the redshift for some values of the parameter $\alpha$. Regardless of the value of $\alpha$, $d_L^{gw}/d_L^{em} \rightarrow 1$ at small $z$, while at higher $z$ goes to a constant. The constant assumes values  larger than 1 for $\alpha>0$ and smaller than 1 for $\alpha<0$. The fact that the ratio goes to $1$ at small $z$ is expected since there cannot be any effect from MG when the source is at $z\simeq 0$.  Moreover when $\alpha<0$ the ratio $d_L^{gw}/d_L^{em}$ is smaller than $1$ thus the source of GWs is magnified and can be seen to larger distances. The opposite holds for $\alpha>0$.

In  Ref.~\cite{Belgacem:2018lbp} it has been proposed the following  phenomenological parametrization:  $d_L^{gw}/d_L^{em}=\Xi(z)$,  where 
\be\label{parametrization}
\Xi=\Xi_0+\frac{1-\Xi_0}{(1+z)^n}\,,
\ee
with $\{\Xi_0,n\}$ being constant free parameters.  This parametrization smoothly interpolates the limits $\Xi(z\ll 1)=1$ and $\Xi(z\gg 1)=\Xi_0$.   The combination of LISA standard sirens with CMB, BAO and SNIa datasets  and the joint analysis of ET standard sirens with the same cosmological dataset allowed to obtain forecasted constraints on the $\Xi_0$ parameter to the percent level accuracy, respectively $\Delta\Xi_0=0.044$~\cite{Belgacem:2019pkk} and $\Delta\Xi_0=0.008$~\cite{Belgacem:2018lbp}, while the uncertainty on $n$ remains large. 

We make a correspondence between  the $f(Q)$-gravity and the phenomenological parametrization in Eq.~\ref{parametrization} by identifying:
\be
\Xi_0\simeq \frac{1}{2}(1+f_{Q0}) \,, \qquad n\simeq \l( \frac{f^\prime_Q}{f_Q-1}\r)_0\,,
\ee
where $f_{Q0}\equiv f_Q(z=0)$ and prime is the derivative with respect to $\ln a$. For the model in Eq.~\eqref{model}, we find 
\be
\Xi_0\simeq1+\frac{\alpha}{4\sqrt{6}}\,, \qquad  n\simeq\frac{1}{2}(3\Omega_{\rm m}^0+4\Omega_{\rm r}^0)\,,
\ee
where $\Omega_{\rm m}^0$ and $\Omega_{\rm r}^0$ are the present time values of the density parameters of matter and radiation components respectively.
Considering the 1$\sigma$ forecasted error on $\Xi_0$ and its $\Lambda$CDM fiducial ($\Xi_0=1$) we  infer  for the derived parameter $\alpha$ ($\alpha_{\rm fiducial}=0$), the following  forecasted errors: $\Delta \alpha \simeq 0.078$ for ET and $\Delta \alpha \simeq 0.43$ for LISA standard sirens.

\section{Conclusion}

We have presented theoretical predictions on linear cosmological observables from a modified gravity model based on the non-metricity scalar, $Q$.  For this class of models a general function of $Q$, $f(Q)$, is included in the action.  The first derivative of $f$ with respect to $Q$ is identified with a time dependent Planck mass and constitutes the source of the modification with respect to the  $\Lambda$CDM behavior at large linear scales. The specific model we analyzed in this work does not change the expansion history with respect to $\Lambda$CDM, thus we focused on the scalar angular power spectra and matter power spectrum as well as on the GWs propagation. 

The $f(Q)$-gravity model studied in this paper is given in Eq.~\eqref{model} and it has one extra free parameter, $\alpha$. Depending on the sign of $\alpha$ we found measurable and specific signatures. The anisotropic stress parameter is equal to 1, and from this it follows that  the effective gravitational coupling is equal to the light deflection parameter.
 In detail, we found that values of  $\mu>1$ ($\alpha<0$) enhances both the matter power spectrum and the lensing potential auto-correlation power spectrum in comparison to the $\Lambda$CDM  model.  In turns, modifications in the lensing potential impact on the low-$\ell$ ISW tail of the CMB TT power spectrum due to a modified late-time ISW effect, which for $\alpha<0$ is suppressed with respect to the $\Lambda$CDM model. This aspect revealed to be the key feature for a better fit to data compared to the $\Lambda$CDM scenario in other MG models.   The case  $0<\mu<1$ ($\alpha>0$) generates a suppressed lensing power spectrum and an enhanced low-$\ell$ CMB TT power spectrum.  Additionally it allows to  realize weaker gravity than $\Lambda$CDM corresponding to a suppressed matter power spectrum. As discussed in Section~\ref{Sec:effects}, these features need to be tested against data, as some of them are very promising in light of   the $\sigma_8$ tension  arising from the mismatch at more than 4$\sigma$ in the measurements by Planck  and that obtained from WL observations, and in the interpretation of the lensing excess in the CMB  Planck temperature data. The model does not show these features at the same time as they are driven by different sign of the $\alpha$ parameter,  thus only a thorough MCMC analysis involving several datasets can give us indications of which one is preferred by data.

Furthermore, the model under investigation includes a modified friction term in the equation of propagation of GWs which introduces a modification of the luminosity distance of standard sirens. We analyzed the predictions of the ratio of GWs luminosity distance and electromagnetic one as  function of the redshift and this showed to follow the phenomenological parametrization introduced in~\cite{Belgacem:2018lbp} in terms of two parameters $\{\Xi_0,n\}$. From the forecasts obtained from standard sirens at ET and LISA  using  this parametrization we computed the relation between $\Xi_0$ and the free parameter of our model and we  deduced the accuracy on the $\alpha$ parameter.  Next generation of GWs detectors will strongly help in constraining deviations due to a running Planck mass and for the case under analysis it will help constarining the $\alpha$ parameter with high accuracy.  

In conclusion, the model shows very interesting signatures which deserve to be tested extensively against data. Additionally   a model selection analysis would provide the information whether the model is statistical preferred by data over the $\Lambda$CDM scenario.  We leave these investigations for future work.

\begin{acknowledgments}
The author thanks B.~J.~Barros,  F.~Pace and D.~Vernieri for useful discussions and comments.
This work is  supported by Funda\c{c}\~{a}o para a  Ci\^{e}ncia e a Tecnologia (FCT) through the research grants UID/FIS/04434/2019, UIDB/04434/2020 and UIDP/04434/2020, by FCT project ``DarkRipple -- Spacetime ripples in the dark gravitational Universe" with ref.~number PTDC/FIS-OUT/29048/2017 and  FCT project ``CosmoTests -- Cosmological tests of gravity theories beyond General Relativity" with ref.~number CEECIND/00017/2018.

\end{acknowledgments}

\appendix


\bibliographystyle{aipnum4-1}
\bibliography{biblio}

\end{document}